\documentclass[]{aa}
\usepackage{graphics,latexsym,amssymb,times,psfig}

\def\gsim{ \lower .75ex \hbox{$\sim$} \llap{\raise .27ex \hbox{$>$}} } 
\def\lsim{ \lower .75ex\hbox{$\sim$} \llap{\raise .27ex \hbox{$<$}} } 

\begin{document}

\title{Structured jets in TeV BL Lac objects and radiogalaxies.}
\subtitle{Implications on the observed properties}

\author{Gabriele Ghisellini \inst{1}, Fabrizio Tavecchio \inst{1}
and Marco Chiaberge \inst{2}}

\offprints{G. Ghisellini; gabriele@merate.mi.astro.it}
\institute{
INAF -- Osservatorio Astronomico di Brera, via Bianchi 46, 
I--23807 Merate, Italy;
\and IRA/CNR, via Gobetti 101, I--40129, Bologna, Italy
}

\date{Received 2004}
 
\titlerunning{Structured jets}
\authorrunning{G. Ghisellini, F. Tavecchio \& M. Chiaberge}

\abstract{
TeV BL Lacertae objects require extreme relativistic bulk motion 
in the gamma--ray emission region, but at the VLBI scale 
their radio knots hardly move. The same sources show 
evidence, in radio, of a structure made of a fast spine 
plus a slow layer. We propose that this structure exists 
even on the spatial scale of regions responsible for the 
gamma--ray emission.
One component sees the (beamed) 
radiation produced by the other, and this enhances the inverse 
Compton emission of both components. 
In addition, this allows the magnetic field to be nearly in 
equipartition with the emitting particles.
The inverse Compton emission of the spine is anisotropic 
in its frame, possibly producing the deceleration of the 
spine by the Compton rocket effect. 
In this scenario, also the slow layer is a relatively 
strong high energy emitter, and thus radiogalaxies 
become potentially detectable by GLAST.
}
\maketitle

\section{Introduction}

There is growing evidence from VLBI studies that pc scale jets in
strong TeV BL Lacs move slowly (Edwards \& Piner 2002; Piner \&
Edwards 2004; Giroletti et al., 2004).  On the contrary, the bright
and rapidly variable TeV emission implies that at the jet scales where
this emission originates, the jet should be highly relativistic.  This
is necessary in order to avoid the absorption of TeV photons by the IR
radiation produced cospatially to the TeV emission (see e.g. Dondi \&
Ghisellini 1995).  Furthermore, fitting the SED of TeV sources with a
simple, one--zone homogeneous synchrotron self--Compton (SSC) model
allows to determine a unique set of physical parameters of the
emitting region (Tavecchio, Maraschi \& Ghisellini 1998), and in fact
all authors applying SSC models derive similar values of the Doppler
factor, in the range 10--20, when the TeV spectrum is not de--reddened
by the absorption of TeV photons by the IR background (Tavecchio et
al. 2001, Kino, Takahara \& Kusunose 2002; Ghisellini, Celotti \&
Costamante 2002; Katarzynski, Sol \& Kus 2003), and larger (up to 50)
when the TeV spectrum is dereddened (Krawczynski, Coppi \& Aharonian
2002; Konopelko et al. 2003).  It is therefore clear that the jet must
suffer a severe deceleration from the $\gamma$--ray emitting zone
($\sim$0.1 pc from the jet apex) to the VLBI ($\sim$1 pc) scale.

Prompted by these observational evidences, Georga\-no\-poulos \&
Kazanas (2003, hereafter GK03) have proposed that if the entire jet is
rapidly decelerating in the $\gamma$--ray zone, then the base of the
jet, still moving fast, will see the radiation produced at the end of
the deceleration zone relativistically boosted.  This ``extra"
radiation will favor the inverse Compton emission, allowing to derive
less extreme values of the physical parameters with respect to a pure
one--zone SSC model.  Our paper is germane to the one of GK03, but we
study the alternative hypothesis that the jet is structured not in the
radial direction, but in the transverse one, being composed by a slow
layer and a fast spine.  

We are motivated by the recent observational evidence coming from
detailed VLBI (including space VSOP observations) radio maps, showing,
in Mkn 501, a {\it limb brightening} morphology, interpreted as
evidence of a slower external flow surrounding a faster spine
(Giroletti et al. 2004).  Similar results have been obtained for a few
radiogalaxies (Swain et al. 1998; Owen et al. 1989; Giovannini et
al. 1999).  Apart from observational evidence, a spine--layer
configuration for the jet has been proposed in the past on the basis
of theoretical arguments (e.g., Henri \& Pelletier 1991). In addition,
the existence of a velocity structure has also been suggested to
explain some observed properties of radiogalaxies, such as their
magnetic field configuration (Komissarov 1990; Laing 1993), and to
overcome problems unifying radiogalaxies with BL Lac objects
(Chiaberge et al. 2000).

We also hope that this assumption helps to find a possible reason for
the deceleration of (at least a part of) the jet, which was {\it
postulated} by GK03.  A slow layer could in fact be the result of the
interaction of the ``walls" of the jet with the ambient medium, or
simply be the result of a jet acceleration which is a function of the
angular distance from the jet axis,
\footnote{Note that the possibility of structured jets
has been proposed and somewhat explored also for Gamma--Ray Bursts,
see Rossi, Lazzati \& Rees 2001; Zhang \& Meszaros 2002.}
producing a velocity structure.
Kelvin--Helmoltz instabilities (for review see Ferrari 1998), 
while important for the formation of the layer, may not decelerate 
(and destroy) the entire jet (e.g. Bodo et al. 2003), 
especially if the jet itself is 
not continuous, but ``intermittent", as in the internal shock 
scenario (Ghisellini 1999, Spada et al. 2001, Guetta et al. 2004).
Note in fact that the dynamical timescale involved in internal shocks
is approximately the light crossing time across the source,
while instabilities can grow with the sound speed.

As in the decelerating scenario proposed by GK03,
there will be a strong radiative interplay and feedback between
the layer and the spine: both parts see extra seed photons coming
from the other part, and this will enhance the inverse Compton
emission of {\it both} components.
This may help explaining why also radiogalaxies can be relatively strong
$\gamma$--ray emitters, as suggested by the Comptel and EGRET
(onboard the Compton Gamma Ray Observatory)
detection of Centaurus A (Steinle et al. 1998 and references therein), 
the recent identifications of NGC 6251 with an EGRET source 
(Mukherjee et al. 2002),
and the possible detection of M87 at TeV energies (Aharonian et al. 2003).
This emission, coming from the inner part of the relativistic jet
of radiogalaxies,
should be characterized by a pronounced variability: this could
help to distinguish it from the high energy radiation coming
from the more extended (kpc) parts of the jet, as suggested by
Stawarz et al. (2003).

In Section 2 we present the basic assumptions of the model and in
Section 3 we discuss what we think is a major outcome of the scenario
we are proposing, namely the possibility that the spine can recoil
under the effect of its own inverse Compton emission (Compton rocket
effect). We apply the model in Section 4 to Mkn 421, Mkn 501, Cen A
and also to NGC 6251, assumed to be ``paired" with the classic BL Lac
PKS 0735+178.  Even if the presented ``fits" 
{\footnote{We call ``fit" what really is a
comparison of the model against the observational data.}
}
are not unique solutions
(given the large number of free parameters), they illustrate the
radiative feedback between the spine and the layer.  Since it is
natural, in our model, that also the inverse Compton emission of the
layer is enhanced, we stress in Section 5 that radiogalaxies, not only
blazars, can be bright $\gamma$--ray emitters, and we present a
preliminary list of possible candidates for detection by GLAST.  In
Section 6 we discuss our results and draw our conclusions.

\section{The model}

\begin{figure}
\psfig{figure=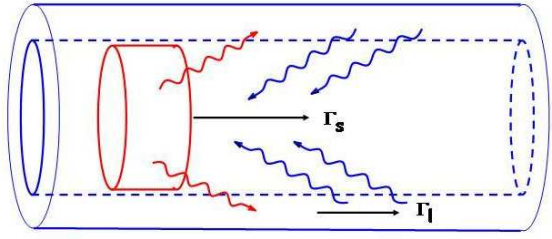,angle=0,width=8cm}
\caption{Cartoon illustrating the layer+spine system.
}
\label{cartoon}
\end{figure}

We assume that the layer can be approximated as a hollow
cylinder,
with external radius $R_2$, internal radius $R$ and width $\Delta R_l^"$, 
as measured in the comoving frame of the layer
\footnote{
Primed quantities are in the rest frame of the spine,
double primed quantities are in the frame of the layer.}.
The comoving volume of the layer is then 
$V^"_l=\pi (R_2^2-R^2)\Delta R_l^"$.
Also for the spine we assume a cylindrical geometry, with the same $R$ 
and a width $\Delta R_s'$, as measured in the comoving frame of the
spine. 
The active volume of the spine is then $V^{'}_s=\pi R^2 \Delta R_s'$.
Fig. \ref{cartoon} illustrates the assumed geometry.
The Lorentz factors of the spine and of the layer are
$\Gamma_s$ and $\Gamma_l$, respectively, with $c\beta_s$ and
$c\beta_l$ the corresponding velocities.
Since the spine and the layer move with different Lorentz factors,
the radiation emitted by the spine (layer) is seen boosted by the
layer (spine).
With respect to a comoving observer at the same distance from the spine
(layer), the radiation energy density is enhanced by a factor 
$\sim (\Gamma^\prime)^2$, with $\Gamma^\prime$ given by
\begin{equation}
\Gamma^\prime\, =\, \Gamma_s\Gamma_l(1-\beta_s\beta_l)
\label{grel}
\end{equation}
Both structures emit by the synchrotron and the inverse Compton processes.
The energy distribution of the emitting electrons, $N(\gamma)$,
is assumed to extend down to $\gamma_{\rm min}$,
and is assumed to have the shape:
\begin{eqnarray}
N(\gamma)&=& K \gamma^{-n_1} \left[ 1+ 
\left({\gamma\over \gamma_b}\right)^{n_1-n_2}\right]
e^{-{\gamma\over\gamma_{\rm cut}}};\quad \gamma>\gamma_{\rm min}
\nonumber \\
N(\gamma)&=&0, \quad\quad\quad\quad\quad\quad\quad\quad\quad\quad\quad
\quad\quad\quad\,\,\,
 \gamma\le\gamma_{\rm min}
\label{ng}
\end{eqnarray}
The normalization (i.e. $K$) of this distribution is
found by imposing that $N(\gamma)$ produces a given
intrinsic synchrotron luminosity, which is an input
parameter of the model.

The seed photons relevant for the scattering process are produced
not only by the spine (layer) electrons, but also by the layer (spine) ones.
There is a strong {\it feedback} between the two components, which
determines the amount of inverse Compton radiation emitted 
by both structures
(if this process is important with respect to the synchrotron one).
As a general rule, this feedback increases the inverse Compton flux, since
both the spine and the layer see an enhanced radiation energy density.
Since the ratio between the radiation and the magnetic energy densities
$U_{\rm B}/U_{\rm rad}$
measures (even if it is not equal, for scatterings in the Klein 
Nishina regime)
the ratio between the inverse Compton and the synchrotron luminosities,
an enhanced $U_{\rm rad}$ in turn implies a larger magnetic field.
Therefore both the synchrotron and the inverse Compton luminosities
can be produced by a reduced number of relativistic electrons
(radiatively cooling in a shorter time with respect to the case
of no feedback).
This fact bears important consequences for the energetics and 
the dynamics of the jet, and it will be discussed in Section 3,
while in Section 4 we give an illustrative example for the 
case of Mkn 421.

The length of the layer, as observed in a frame comoving with the
spine, is $R^\prime_l = R^"_l/\Gamma^\prime$.
Analogously, the length of the spine, as observed in the frame
comoving with the layer, is $R^{"}_s = R^\prime_s/\Gamma^\prime$.
In the following, we will always assume that the layer is longer
than the spine, even in the frame of the spine.
To calculate the radiation energy density of one component
as observed by the other, we will assume the following:
\begin{itemize}
\item Let us call $\bar R\equiv (R_2+R)/2$.
In the comoving frame of the layer,
the radiation energy density
in the entire cylinder is assumed to be $U^"_l = L^"_l /(\pi \bar R^2 c)$.
In the frame of the spine, this radiation energy density
is assumed to be boosted by a factor $(\Gamma^\prime)^2$, i.e.
$U^\prime_l = (\Gamma^\prime)^2 U^"_l$.
\item In the comoving frame of the spine,
the radiation energy density within the spine is assumed to 
be $U^\prime_s = L^\prime_s /(\pi R^2 c)$.
In the frame of the layer, this radiation energy density
is observed to be boosted by $(\Gamma^\prime)^2$,
but also diluted (since the layer is larger than the spine)
by the factor 
$\Delta R^"_s/\Delta R^"_l=(\Delta R^\prime_s/\Gamma^\prime)/\Delta R^"_l$.
\end{itemize}

While the synchrotron and the SSC emission in the comoving frame
is assumed to be isotropic, the inverse Compton process
between electrons of the layer (spine) and seed photons
produced by the spine (layer) is highly anisotropic.
Dermer (1995) found the pattern
of the emitted radiation from a moving blob immersed in a
bath of seed photons (e.g. corresponding to the radiation 
produced by the broad line region of a powerful blazar),
and pointed out the fact that in this case
the external Compton radiation is more beamed than the synchrotron
and SSC emission.

In our case the component contributing to the ``external" radiation
is not at rest with the distant observer, but moves.
To find out the pattern of the emitted radiation
it is convenient to move to the comoving frame of the emitter
of the seed photons.
Consider then the seed photons produced by the layer,
and an observer comoving with the layer.
In this frame, the spine is moving with $\Gamma^\prime$,
and the photon frequencies produced by the spine are blueshifted by the
Doppler factor 
\footnote{$\delta=\Gamma^{-1}(1-\beta\cos\theta)^{-1}$,
where $\theta$ is the viewing angle.
Note that $\Gamma$ and $\theta$ are different in different frames.
$\delta_{s,l}$ is defined as the beaming factor of the radiation
produced in the spine as observed in the layer.
}
$\delta_{s,l}$.
Going to the frame of the distant observer, these photons
are further blushifted by $\delta_l$.
But the distant observer will see the same photons blushifted
by $\delta_s$.
This implies
\begin{equation}
\delta_{s,l}\, \delta_l \, =\, \delta_s
\end{equation}
We can repeat the same argument for photons produced by the layer.
We then have
\begin{equation}
\delta_{s,l}\, =\, {\delta_s\over \delta_l} \, =\, {1\over \delta_{l,s}}
\end{equation}
This nice argument is due to GK03.

In the frame of the layer the external Compton radiation produced
by the spine follows a pattern  $\propto \delta_{s,l}^{4+2\alpha}$
(Dermer 1995),
where $\alpha$ is the spectral index of the emission 
[$F(\nu) \propto \nu^{-\alpha}]$,
while the synchrotron and SSC emission follows the usual pattern
$\propto \delta_{s,l}^{3+\alpha}$.
If $I^\prime(\nu^\prime)$ is the monochromatic intrinsic intensity
produced by the spine, we have
\begin{eqnarray}
I(\nu) \, &=&\, I^\prime(\nu^\prime) \delta_{s,l}^{4+2\alpha}\delta_l^{3+\alpha} 
\, =\,  I^\prime(\nu^\prime)\delta_s^{3+\alpha}\,
\left({\delta_s\over \delta_l}\right)^{1+\alpha} {\rm (EC)} 
\nonumber \\
I(\nu) \, &=&\, I^\prime(\nu^\prime) \delta_s^{3+\alpha}  \quad
{\rm (S,~SSC)}
\end{eqnarray}
Here S stands for synchrotron, EC for ``external Compton"
(scattering of the seed photons coming from the layer). 
In the case of radiation produced by the layer, the transformation
is the same, with $\delta_s$ and $\delta_l$ interchanged.

\vskip 0.5 true cm
\noindent
{\bf Free parameters ---}
For each component the input parameters are:  
$\Delta R'$, $B$, $L'_{\rm inj}$,
$\Gamma$, $\gamma_{\rm min}$, $\gamma_{\rm b}$, 
$\gamma_{\rm cut}$, $n_1$,$n_2$.
Parameters equal for both components are $R$ and the 
viewing angle $\theta$. 
The outer radius of the layer $R_2$ must also be specified,
but we will always use $R_2=1.2 R$.
For $n_2>3$, $\gamma_{\rm cut}$ becomes unimportant and we have
a total of 18 parameters.
While they appear (and are) many, we stress that the aim of
this paper is to discuss the main effects of having the layer+spine
structure, and not (yet) the exact determination of the physical
quantities inside the source.
In other words, we have to model {\it two} structures
by observing the radiation coming from only one of them,
and this will leave some ambiguity, unless we find information
about the layer (spine) even if we are observing the
spine (layer) emission
\footnote{One possibility could be the observation, in radiogalaxies,
of emission lines resulting from photoionization due to the spine radiation
(Morganti et al. 1992), 
or the knowledge of the total power carried by the entire jet
through estimates of the total energy and age of the extended radio structure
(as done by Rawlings \& Saunders 1991).}.

\section{Energetics and dynamics}

As we will see in the next Section, the power spent by the jet to emit
the inverse Compton radiation is significant, if compared with the
total bulk kinetic power contained in the relativistic electrons and
protons.  As described above, this luminosity is emitted
anisotropically in the comoving frame of the spine, hence the spine
must recoil. Since this {\it Compton rocket} effect is a potentially
important deceleration mechanism, we discuss it here in more detail.

The most convenient frame where to study the dynamics of the spine
is the frame comoving with the layer.
In this frame the synchrotron radiation produced by the
layer is isotropic, with energy density $U^"_{\rm syn}$.
To avoid an excess of symbolism, 
from now on (unless otherwise noted) the unprimed
quantities are measured in the frame of the layer,
and the primed quantities are evaluated in the frame
of the spine.
For the same reason we let the random Lorentz factor
unprimed, with the notion that these Lorentz factors
are measured in the frame of the spine, where they are
assumed to be isotropically distributed.

The total Lorentz factor ($\tilde\gamma$) of the electrons is the
superposition of the relative bulk ($\Gamma$) and random ($\gamma$)
Lorentz factors.  Assume that the random velocity forms an angle
$\theta^\prime$ with respect to the jet axis, 
in the spine frame. From Rybicki \& Lightman (1979) we have:
\begin{eqnarray}
\beta_x \, &=& \, { \beta^\prime\cos\theta^\prime +\beta_{bulk} 
\over 1+\beta_{bulk}\beta^\prime\cos\theta^\prime} \nonumber \\
\beta_y \, &=& \,{ \beta^\prime\sin\theta^\prime  \over 
\Gamma_{bulk}(1+\beta_{bulk}\beta^\prime\cos\theta^\prime )}
\end{eqnarray}
The total $\tilde\gamma^2$ is
\begin{eqnarray}
\tilde\gamma^2 \,  &=& \, \left[ 1-\beta_x^2-\beta_y^2\right]^{-1} \nonumber \\
&=&\,
(1+\beta_{bulk}\beta^\prime\cos\theta^\prime )^2 \gamma^2\Gamma^2 
\end{eqnarray}
If the particle distribution is isotropic in the spine frame,
the average over angles gives
\begin{eqnarray}
\langle \tilde\gamma^2\rangle \,  &=&\, 
{ \int 2\pi \sin\theta^\prime \tilde\gamma^2(\theta^\prime) 
d\theta^\prime \over 4\pi } \nonumber \\ 
\, &=&\, {\gamma^2 \Gamma^2\over 2} \, \int_{-1}^1 
(1+\beta_{bulk}\beta^\prime\mu^\prime)^2 d\mu^\prime 
\nonumber\\ \, &=&\, 
\left[ 1+{(\beta_{bulk}\beta^\prime)^2 \over 3}\right] \gamma^2\Gamma^2
\end{eqnarray}
the factor in the square parenthesis becomes
$(4/3)$ for ultrarelativistic speeds.

The loss of energy of the jet is proportional to $\langle \tilde
\gamma^2\rangle$, and the loss of momentum is the component along the
jet axis of the loss of energy.  In other words, we have to calculate
the quantity
%
%

%
\begin{equation}
\langle \tilde \gamma^2\rangle_{z}\, =\, 
{ \int 2\pi \mu^\prime \tilde\gamma^2(\theta^\prime) 
d\mu^\prime \over 4\pi } \, =\, 
{2\over 3}\, \beta_{bulk}\beta^\prime \gamma^2\Gamma^2
\end{equation}

Now assume that the spine carries $N_p$ protons (total number)
and $N_e$ leptons. 
The loss of momentum of the jet is described by:
\begin{eqnarray}
{d\Gamma \over dt} \, 
&=&\, {4\over 3}  {\sigma_T c N_e U_{\rm syn} \langle \tilde \gamma^2\rangle_{z}
\over N_p m_p c^2 + N_e \langle\gamma\rangle m_ec^2} \, 
\nonumber \\
&=&\, {8\over 9} {\sigma_T c N_e U_{\rm syn} \langle \gamma^2\rangle \Gamma^2
\over N_p m_p c^2 + N_e \langle\gamma\rangle m_ec^2}  \,
\label{dec} 
\end{eqnarray}
%
%
We remind that the bulk Lorentz in Eq. (\ref{dec}) is the relative 
Lorentz factor 
(which is the Lorentz factor of the spine as measured in the layer).
A detailed analysis of this Compton rocket effect in  AGN jets has
been presented by Sikora et al. (1996). They derive the equations
regulating the drag effect and apply the results to the case of jets in 
radio--loud quasars with strong emission lines. 
Eq. (\ref{dec}), although obtained in a
rather simplified way, coincides with the results of Sikora et
al. (for the specific case of isotropic radiation field). 

To find the value of $\Gamma_s$ as measured by an observer at earth,
one must invert Eq. \ref{grel}.
It must be noted that, in general, $\langle\gamma\rangle$ and 
$\langle\gamma^2\rangle$
are not constant, but can change due to e.g. radiative cooling.
We have integrated Eq. \ref{dec} assuming:
\begin{itemize}
\item 
The particle distribution remains unaltered for a time 
equal to the light crossing time of the spine (i.e. $\Delta R_s/c$)
as measured in the frame of the layer.
This corresponds to assume that the time of the injection of particles
throughout the spine lasts for a similar time.
\item
After this time we calculate the actual $\gamma_{\rm cool}$ at 
the given time, and assume that the particle distribution
vanishes for $\gamma>\gamma_{\rm cool}$. 
We neglect adiabatic losses, for simplicity.
\item
We then calculate the new values of
$\langle\gamma\rangle$ and $\langle\gamma^2\rangle$
and the new value of $\Gamma$.
\item
We invert Eq.\ref{grel} and finally find $\Gamma_s$.
\end{itemize}

\begin{figure}
\vskip -0.5 true cm
\psfig{figure=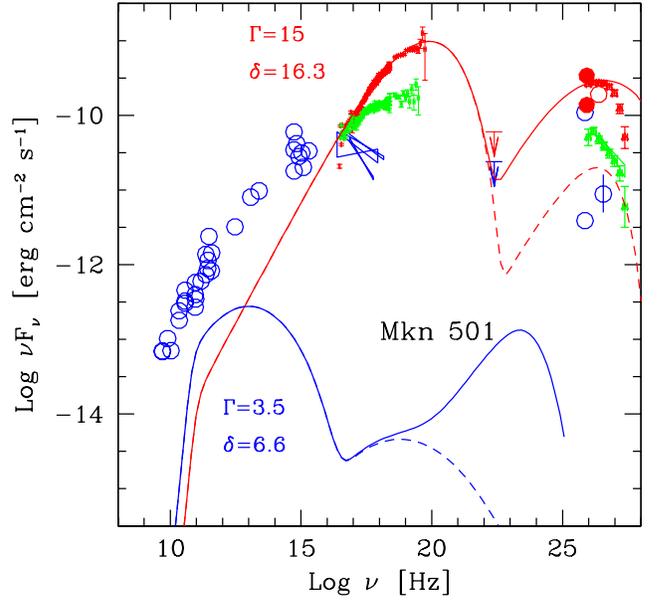,angle=0,width=9.5cm}
\vskip -0.5 true cm
\caption{Example of the SED produced by the spine--layer
system, using the parameters listed in Tab. 1.
Dashed lines correspond to the emission
of the spine (layer) without taking into account the seed 
photons coming from the layer (spine). Data from Pian et al. (1998)
and Djannati-Atai et al. (1999).
}
\label{501}
\end{figure}

\section{Results}

\begin{figure}
\vskip -0.5 true cm
\psfig{figure=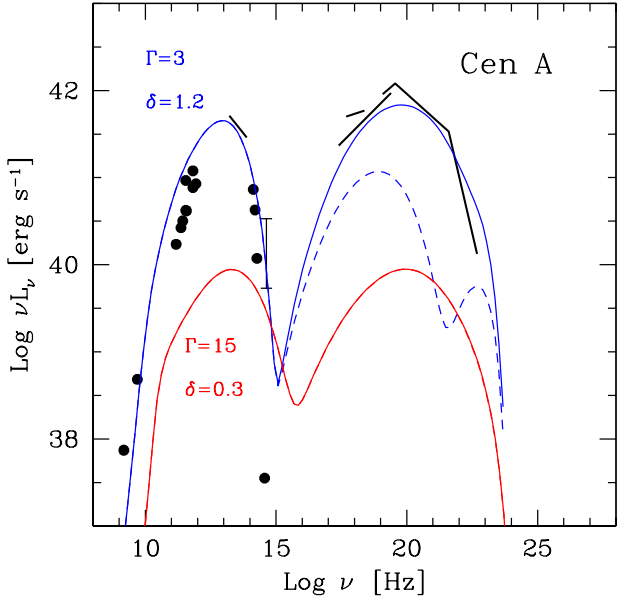,angle=0,width=9.5cm}
\vskip -0.5 true cm
\caption{The SED of Cen A is modelled by the spine--layer
system, using the parameters listed in Tab. 1.
Dashed lines correspond to the emission
of the spine (layer) without taking into account the seed 
photons coming from the layer (spine).
For the spine the dashed and continuous lines
overlap. Data from Chiaberge, Capetti \& Celotti (2001) and
references therein.
}
\label{cena}
\end{figure}

\subsection{Fitting Mkn 501, Mkn 421 and Cen A}

We here apply the model to the two best studied 
TeV BL Lacs: Mkn 501 and Mkn 421, and to the radiogalaxy Cen A.


We stress that our aim is not to precisely fit the SED
of these sources, since our results are not unique.
For instance, the parameters for the layer of Mkn 421 and Mkn 501
have been chosen with the aim to increase the allowed value
of the magnetic field of the corresponding spines.
In other words, we applied a theoretical prejudice, aiming to bring
these sources closer to equipartition than a simple
one zone SSC model allows to.
Other poorly constrained parameters are the width of the layer
and its bulk Lorentz factor.
For the former we simply require $\Delta R''>R$, for the latter
we have the requirement that the counterjet is, in these sources,
invisible up to the few tens of m.a.s. scale, leading to a limit
of $\Gamma_l>2$--3.

In conclusion, our aim is to demonstrate that our model
can consistently work both for radiogalaxies and BL Lacs, and
this allows to draw some general conclusions about their
jets and the importance of the inverse Compton emission.
The input parameters used for the models shown in Figs. 2--5
are all listed in Table 1, while Table 2 reports the values
of the kinetic powers carried by the spine and the layer,
the average (and the average of the square) of the random 
Lorentz factor of the emitting electrons, and the relative
bulk Lorentz factor $\Gamma^\prime$.
The kinetic powers are defined as
\begin{equation}
L_{i} \, =\, \pi R^2 \Gamma^2 \beta c U_{i}
\end{equation}
where the energy density $U_i$ refers to:
electrons ($U_e = \int N(\gamma)\gamma m_ec^2 d\gamma$);
magnetic field ($U_B=B^2/8\pi$);
protons ($U_p$), assuming one cold proton per electron;
total radiation ($U_{\rm rad}$) and
synchrotron radiation ($U_{\rm syn}$).

In all cases shown in Figs. 2--5, continuous lines show 
the results of the models including the spine--layer 
radiative feedback, while the dashed lines are the 
results of the same models neglecting the feedback.

\begin{figure}
\vskip -0.5 true cm
\psfig{figure=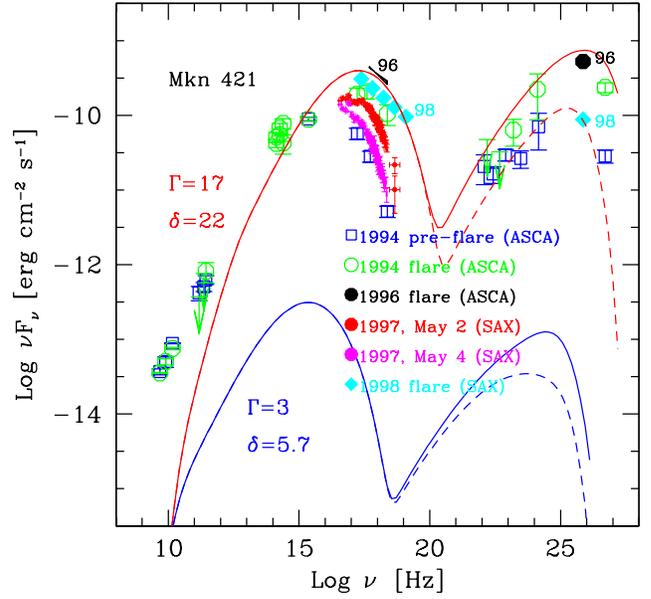,angle=0,width=9.5cm}
\vskip -0.5 true cm
\caption{Example of the SED produced by the spine--layer
system, using the parameters listed in Tab. 1.
Dashed lines correspond to the emission
of the spine (layer) without taking into account the seed 
photons coming from the layer (spine).
For the data points see Costamante \& Ghisellini (2002)
and references therein.
}
\label{421}
\end{figure}

\begin{itemize}

\item
{\bf Mkn 501:} 
Fig. 2 shows the ``fits" to the SED of Mkn 501 in its 
flaring state of April 1997 (Pian et al. 1998).
Note the rather moderate bulk Lorentz factor
used for the spine ($\Gamma_s=15$), which has the same
value of the one used by GK03.
The rather large value of the magnetic field ensures nearly
equipartition between magnetic and particle energy.

\item
{\bf Mkn 421:} Fig. 3 shows our model compared with
different simultaneous SED of this source, and aiming
to ``fit" the highest state.
The bulk Lorentz factor of the spine is $\Gamma_s=17$,
which at a viewing angle $\theta=2.5^\circ$ leads to
a beaming factor $\delta=22$.
The remarkable difference with respect to a simple
one zone SSC models is the much larger value of the 
magnetic field: 1 Gauss compared with $B\sim 0.02$--0.1 Gauss
(see the references quoted in the Introduction).
This brings the magnetic energy in equipartition with the particle
energy, dominated in this case by the energetic electrons,
which have a very large average random Lorentz factor 
($\langle\gamma\rangle=1900$,
and therefore $\langle\gamma\rangle m_e \sim m_p$).
This is due both to the relatively large value of $\gamma_{\rm min}$
and the flat electron slope at low energies, in turn required
in order not to overproduce the IR emission.
\item
{\bf Cen A:} As can be seen in Fig. \ref{cena}, the synchrotron 
spectral component in this object is particularly narrow, with a peak
in the far IR.  This indicates a corresponding narrow energy
distribution of the emitting electrons.  We have assumed a viewing
angle of $\theta=40^\circ$, corresponding to a modest beaming of the
layer emission ($\delta=1.2$), and a severe de--beaming of the spine
radiation ($\delta=0.28$).  Comparing the jet powers of this source
with Mkn 421 or Mkn 501, we find that the jet of Cen A is more
powerful, with most of the power carried by the spine.

\end{itemize}

\subsection{``Pairing" BL Lacs with radiogalaxies}

We here try to see if our spine+layer 
structure can explain {\it at the same time} the emission
from a ``classical" BL Lac object,
PKS 0735+178, and the radiation observed for
the radiogalaxy NGC 6251, which has been recently associated
with an EGRET source.
In other words, we check if these two apparently
very different objects, at very different distances 
(but both emitting high energy $\gamma$--rays), can be 
considered as ``paired". 
Note that the radio luminosity $L_{1.4}$ at 1.4 GHz of the two objects,
indicative of the extended (and unbeamed) radio power, is almost equal 
(for PKS 0735+178: $L_{1.4}=9\times 10^{31}$ erg s$^{-1}$ Hz$^{-1}$, 
Cassaro et al. 1999;  
for NGC 6251: $L_{1.4}=3\times 10^{31}$ erg s$^{-1}$ Hz$^{-1}$,
Laing, Riley \& Longair, 1983, extrapolating from 178 MHz with a 
radio spectral index $\alpha=0.7$).
In this case the differences in the observed nuclear SED are due
entirely to a different viewing angle $\theta$, enhancing the spine
or the layer emission for small and large $\theta$, respectively.
As a consequence, we have now much less freedom than before
for the fitting,
but, of course, now 
it is the choice of these two sources which is somewhat
arbitrary, even if both are $\gamma$--ray emitters and presumably
both belong to the FR I class of radiogalaxies.

\begin{figure}
\vskip -0.3 true cm
\psfig{figure=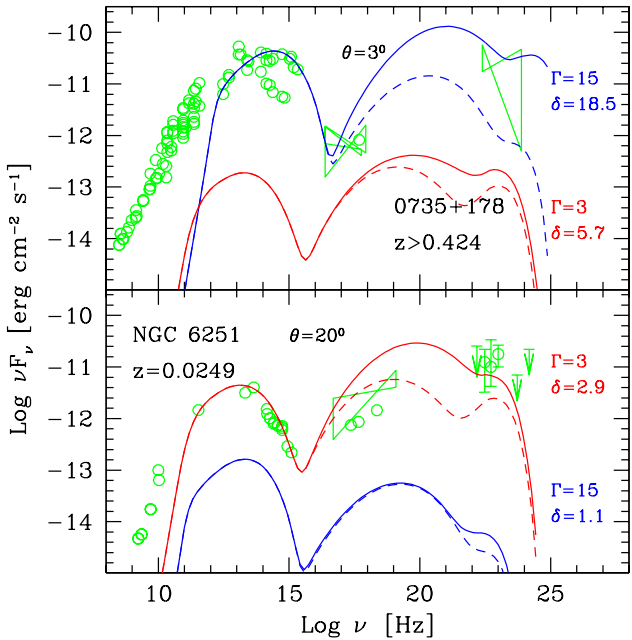,angle=0,width=9.5cm}
\vskip -0.5 true cm
\caption{Example of the SED produced by the spine--layer
system, using the parameters listed in Tab. 1.
The viewing angle is $\theta=2^\circ$
for the example shown in the upper panel,
and $\theta=22^\circ$ for the bottom panel.
The model is compared with the data of the BL Lac
PKS 0735+178 (top) and the FR 1 radiogalaxy NGC 6251
(bottom).
Dashed lines correspond to the emission
of the spine (layer) without taking into account the seed 
photons coming from the layer (spine).
Data from Ghisellini et al. (1998) and Chiaberge et al. (2003)
and references therein. 
}
\label{6251}
\end{figure}

\begin{itemize}

\item {\bf PKS 0735+178:} 
This ``classic" BL Lac has been detected by EGRET (Hartman et al. 1999),
and shows a relatively flat X--ray spectrum, signature
of the fact that at these energies the inverse Compton
emission is already dominating.
The synchrotron peak frequency lies in the IR--optical
and possibly ``moves" within these bands.
The data shown in Fig. 4 are not simultaneous, and therefore
give a rough idea of the entire SED.
Nevertheless it is clear that this source belongs to the LBL
(Low energy peak BL Lac, see Giommi \& Padovani 1995) category.
There is only a lower limit for its redshift,
($z>0.424$); for the modeling we assume $z=0.424$.
The implied energetics are two orders of magnitude larger than
for the previous sources, if a similar value of the Doppler beaming
factor is appropriate.
Accordingly, we have assumed a larger intrinsic power, and a larger
value of the magnetic field.
This brings the magnetic energy closer to equipartition (with 
respect to a pure SSC model) with the energy contained in the
relativistic electrons.

\item {\bf NGC 6251:} The layer assumed to be in the jet of PKS 0735+178,
if observed at $20^\circ$, becomes the dominant contributor to the SED
and give rise to the spectrum of the nearby radiogalaxy NGC 6251 ($z=0.0249$),
as illustrated in the bottom panel of Fig. 4.
Note that also in this case the data points are not simultaneous,
and can give only a rough idea of the overall (nuclear) SED of this object.
The beamed non--thermal component, being variable with
a relatively large amplitude, could dominate only occasionaly the total
flux, letting in other cases the component produced by the accretion
disk be dominant. 
This could also explain why, in some observations, a relatively strong
fluorescent broad iron line (at 6.4 keV) is visible (Gliozzi et al. 2004).
Bearing in mind these uncertainties in the relative strength of the
beamed vs accretion components, 
the slight overproduction of X--rays predicted by the model
is not (yet) necessarily a failure of the model.
Note also that when the layer is not illuminated by the spine
(and this can happen, since the spine can be constituted by discontinuous
blobs), then the predicted emission corresponds to the dashed line,
which is closer to the X--ray data, but underpredicts the $\gamma$--ray flux.
This may originate two 
different states of the source, and EGRET could detect the object only 
when the layer was illuminated by the spine.
Note that this source showed low amplitude, relatively fast 
($\sim 10^4$ s) variability in hard X--rays (above 0.8 keV),
in agreement with the idea that the non--thermal beamed continuum
rivals the contribution produced by the accretion disk, with
the latter
better visible in the softer X--ray band (Gliozzi et al. 2004).

\end{itemize}

\begin{table*}  
\begin{center}  
\begin{tabular}{|l|llllllllllll|}  
\hline      
& $R$    
& $\Delta R'$  
& $L_{\rm syn}^\prime$  
& $B$  
& $\gamma_{\rm min} $  
& $\gamma_{\rm b} $ 
& $\gamma_{\rm cut}$  
& $n_1$
& $n_2$
& $\Gamma$ 
& $\theta$ 
&\\
&cm  &cm &erg s$^{-1}$  &G & & & & & & & &\\  
\hline  
Mkn 501 (spine)               &3e15 &1e15   &2.5e41  &1.3&1   &3e5 &1e7 &1.7 &2.6 &15 &3.5 &  \\ 
Mkn 501 (layer)               &3e15 &3e16   &3.0e39  &1  &1   &3e2 &3e4 &1.5 &3.5 &3.5&3.5 & \\ 
Mkn 421 (spine)               &3e15 &1e15   &3.0e40  &1.1&2e2 &3e4 &1e6 &1.5 &3.5 &17 &2.5 & \\
Mkn 421 (layer)               &3e15 &3e16   &5.0e39  &0.5&50  &1e4 &3e5 &1.5 &3.5 &3  &2.5 & \\
Cen A (spine)                 &1e16 &1e15   &8.0e42  &4  &50  &3e3 &3e4 &1.5 &4   &15 &40 & \\
Cen A (layer)                 &1e16 &1e17   &1.0e42  &2  &1e2 &4e3 &4e3 &1.0 &4   &3  &40 & \\
PKS 0735+178/NGC 6251 (spine) &5e15 &5e14   &1.0e42  &5  &50  &2e3 &6e3 &1.5 &3   &15 &3/20 &  \\
PKS 0735+178/NGC 6251 (layer) &5e15 &2.5e16 &5.0e41  &1.8&1e2 &2e3 &1e4 &1.6 &4.5 &3  &3/20 &  \\ 
\hline  
\end{tabular}  
\caption{Input parameters of the models for the layer and the spine 
shown in Figg. \ref{501}--\ref{6251}. Note that we have assumed that the viewing angles
for PKS 0735+178 and for NCG 6251 are  $3^\circ$ and $20^\circ$ respectively.
} 
\end{center}  
\end{table*}

\begin{table*}  
\begin{center}  
\begin{tabular}{|l|lllllllll|}  
\hline      
& $L_e$    
& $L_p$  
& $L_B$  
& $L_r$  
& $L_s$
& $<\gamma>$  
& $<\gamma^2>$ 
& $\Gamma^\prime$ 
& $\delta$   \\
&erg s$^{-1}$  &erg s$^{-1}$ &erg s$^{-1}$  &erg s$^{-1}$  &erg s$^{-1}$ & &  & & \\  
\hline  
Mkn 501 (spine)               &3.7e42 &6.7e43 &1.3e43 &6.1e43 &5.8e43 &102  &2.1e7  &2.3  &16.3 \\ 
Mkn 501 (layer)               &1.3e41 &1.7e43 &1.7e41 &5.9e40 &3.9e39 &14   &2.9e3  &2.3  &6.6  \\ 
Mkn 421 (spine)               &1.0e43 &9.7e42 &1.2e43 &1.3e43 &8.9e42 &1.9e3 &3.1e7 &3.0  &21.9 \\
Mkn 421 (layer)               &2.5e40 &8.5e40 &3.2e40 &1.2e41 &5.6e39 &5.4e2 &2.8e6 &3.0  &5.7  \\ 
Cen A (spine)                 &2.7e45 &1.9e46 &1.3e45 &4.6e45 &1.8e45 &2.6e2 &2.5e5 &2.7  &0.28  \\ 
Cen A (layer)                 &7.7e41 &4.1e42 &2.7e41 &5.1e43 &9.6e41 &3.5e2 &2.2e5 &2.7  &1.2  \\ 
PKS 0735+178/NGC 6251 (spine) &8.0e44 &7.4e45 &5.3e44 &4.0e44 &2.2e44 &198  &1e5    &2.7  &18.6 \\
PKS 0735+178/NGC 6251 (layer) &2.3e42 &1.6e43 &5.4e41 &3.9e43 &9.4e41 &271  &1.4e5  &2.7  &2.9 \\ 
\hline  
\end{tabular}  
\caption{Derived jet powers and parameters of the models for the layer and the spine 
shown in Figg. \ref{501} -- \ref{6251}.}
\end{center}  
\end{table*}

\subsection{Jet deceleration by the Compton rocket effect}

\begin{figure}
\vskip -0.6 true cm
\psfig{figure=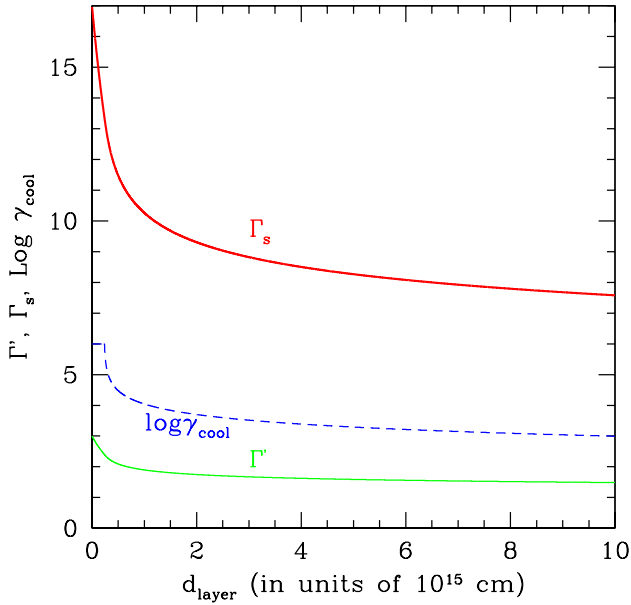,angle=0,width=9.5cm}
\vskip -0.5 true cm
\caption{The bulk Lorentz factor as a function of distance
as measured in the frame of the layer and in the earth frame.
Also shown, as labelled, is the logarithm of $\gamma_{\rm cool}$.
We have adopted the same parameters as the ones used for Mkn 421
and reported in Table 1.}
\label{decel}
\end{figure}

As anticipated above, from Table 2 we can infer that the power
released as inverse Compton radiation by the spines of Mkn 501 and Mkn 421
is comparable to the total
kinetic power carried by the jet.
In this circumstances the {\it Compton rocket} effect is 
important to determine the dynamics.  
For illustration, Fig. \ref{decel} shows the evolution of 
the bulk Lorentz factor in the case of Mkn 421,
calculated along the lines discussed in Section 3,
and assuming the same parameters used for the fitting 
described above and listed in Tab. 1.
We show the
bulk Lorentz factor as measured by the layer and also by the observer
at earth, but both as a function of the distance as measured in the
frame of the layer.  
We also show the logarithm of $\gamma_{\rm cool}$.  
As can be seen, the recoil is very significant, decelerating
the spine from $\Gamma_s=17$ to $\Gamma_s=7.5$ within the layer.  Note
that $\Gamma=3$ is the bulk Lorentz of the layer, which is therefore
the minimum value which the spine could attain.

\subsection{General outcomes and implications}

From the exercise of ``fitting" the SED of the previous few BL Lacs
and radiogalaxies, we can draw the following general conclusions:

\begin{itemize}

\item
The feedback between the layer and the spine enhances the inverse Compton
emission in both components.
In the shown examples, the enhancement is around one order of magnitude,
or even greater.

\item
The magnetic field is larger, in this spine/layer scenario,
and is consistent with being in equipartition with
the relativistic electrons.

\item
The larger magnetic field implies that, to produce the same amount
of synchrotron radiation, less electrons are needed.
If there is one proton per electrons, we need also less
protons. 
This implies a decrease in the total jet kinetic power
(with respect to a pure SSC model).

\item
The Compton rocket effect can influence the dynamics of the
spine, since the external Compton luminosity (being anisotropic
in the comoving frame) becomes comparable with the total
kinetic power carried by the spine itself.

\item 
The examples shown here assume that the spine is active
only when inside the layer.
But other possibilities exist: for instance, the layer
could be particularly narrow (i.e. small $\Delta R_l$), 
and allow the spine to survive the passage through itself
(i.e. the electrons may not cool completely).
In such a case, the spine will continue to emit
the same amount of synchrotron radiation (if the particle
distribution does not change), while the amount of
high energy emission will drastically decrease.
This might explain the ``orphan TeV flares"
(i.e. flares in the TeV band not accompanied by simultaneous
flares in the X--ray band) observed in simultaneous RXTE/TeV 
observations (Krawczynski et al. 2004).

\item
To be effective, the layer must of course be located
in the same region of the jet where the spine undergoes
dissipation and emits.
In other words, it must be compact, even if its width
can be larger (factor $\sim$10) than the spine width
(i.e. $\Delta R^"_l\sim 10 \Delta R^\prime_s$).
This implies variability of the layer.
Even if its particle distribution is steady, in fact, 
its high energy emission will change as 
the spine is illuminating it or not
(e.g. compare the continuous and dashed lines for the layer
in Figg. 2--5).

\end{itemize}

\section{GeV and TeV radiogalaxies?}

\begin{figure}
\vskip -0.5 true cm
\psfig{figure=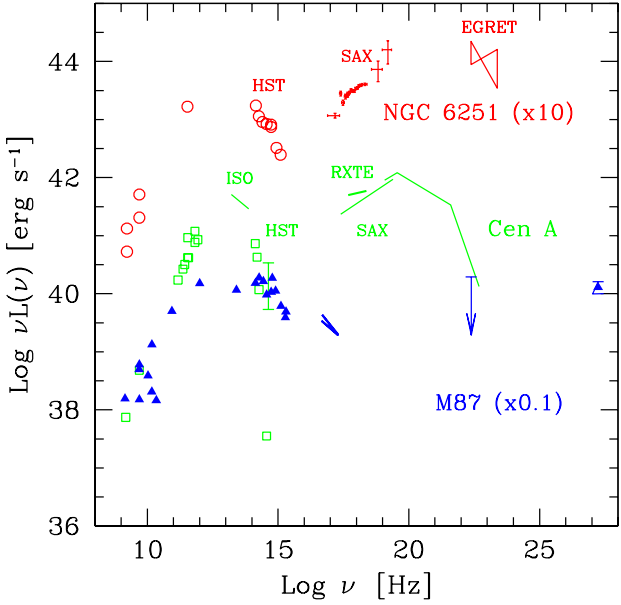,angle=0,width=9.5cm}
\vskip -0.5 true cm
\caption{The SED of NGC 6251, Cen A and M87.
The data of NGC 6251 and M87, for clarity, have been vertically shifted
by the labelled amount.
Note that the SED of these radiogalaxies show the same two--bump 
structure of blazars, as illustrated also in Fig. \ref{sequence}.
Data for M87 are taken from Reimer, Protheroe \& Donea (2004) and reference
therein, except for the X--rays  (Marshall et al. 2002), 
and the EGRET upper limit (Fichtel et al. 1994).
}
\label{all}
\end{figure}

\begin{figure}
\vskip -0.5 true cm
\psfig{figure=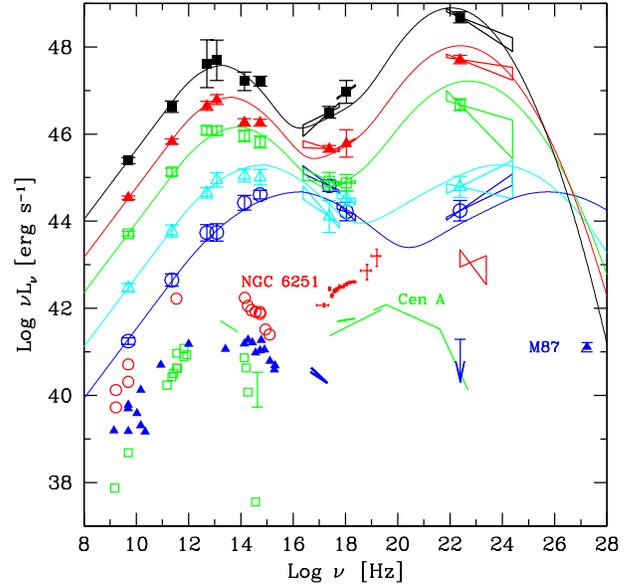,angle=0,width=9.5cm}
\vskip -0.5 true cm
\caption{The SEDs of NGC 6251, Cen A and M87 are compared with the
blazar sequence, as proposed by Fossati et al. (1998).
The hard X--ray [2--10 keV] 
spectra of blazars come from the work of Donato et al. (2001). 
}
\label{sequence}
\end{figure}

Also the slow layer produces a lot of GeV radiation, which 
remains visible even at large viewing angles
(consider that $\theta = 1/\Gamma_{\rm layer}\sim 20^\circ$).
Therefore also radiogalaxies should be high energy emitters.
Indeed, the three radiogalaxies detected at high energies so far, 
shown in Fig. \ref{all}, show the characteristic double bump SED
typical of blazars, suggesting a similar origin of their emission.
This is more clear in Fig. \ref{sequence}, which compares
directly the averaged SED of blazars with the SED of Cen A, NGC 6251
and M 87.

For Cen A and NGC 6251, which have been identified with EGRET surces,
we measure a ratio between the EGRET ($\nu F_\nu$) flux at 100 MeV
and the radio flux of the core at 5 GHz which is 60 and 300,
respectively.
This large ratio is also consistent with the TeV detection of M87,
if the peak of its $\gamma$--ray emission lies between the
EGRET and the TeV band.

If all FR I radiogalaxies have a similar $\gamma$--ray to radio flux
ratio, we can identify the best candidates for detection in 
future AGILE and GLAST observations.
To this aim, we have averaged the ratio of the $\gamma$--ray to radio flux 
for the three radiogalaxies detected in $\gamma$--rays which results in
\begin{equation}
\log \nu_{\gamma}F_{\gamma} \, =\, (2\pm 0.5) \log \nu_{\rm R} F_{\rm R, core}
\label{ratio}
\end{equation}
where $F_{\rm R, core}$ is the radio flux of the core at 5 GHz
(Giovannini et al. 1998; Chiaberge, Capetti \& Celotti 1999 and references therein),
and $F_{\gamma}$ is the $\gamma$--ray flux at 100 MeV. 
Using Eq. \ref{ratio} as an empirical guide to predict the $\gamma$--ray fluxes
of radiogalaxies, we list in
Tab. 3 the FR I radiogalaxies belonging to the 3C sample sample, in order 
of decreasing nuclear radio power, and the corresponding predicted 100 MeV flux.
Assuming a sensitivity limit of GLAST
of $\sim 5\times 10^{-13}$ erg cm$^{-2}$ s$^{-1}$
above 100 MeV for one year of exposure time,
i.e. a factor 20 better than EGRET,
(see e.g.: http://glast.gsfc.nasa.gov),
we have that more than a dozen FR I radiogalaxies can be detected.

\begin{table}
\begin{center}
\begin{tabular}{|l| c c|}
\hline
       Name  &    $\log F_{\rm R, core}$ &  $\log \nu_\gamma F_\gamma$    \\
             &    (5 GHz)               &(100 MeV) \\
             &   erg cm$^{-2}$ s$^{-1}$ Hz$^{-1}$ & erg cm$^{-2}$ s$^{-1}$ \\ 
\hline
          3C~84   &   --21.37 & --~~9.67$\pm0.5$  \\
         3C~274   &   --22.40 & --10.70$\pm0.5$  \\
          3C~78   &   --23.02 & --11.32$\pm0.5$  \\
         3C~317   &   --23.41 & --11.71$\pm0.5$  \\
         3C~270   &   --23.51 & --11.81$\pm0.5$  \\
         3C~465   &   --23.57 & --11.87$\pm0.5$  \\
         3C~346   &   --23.66 & --11.96$\pm0.5$  \\
         3C~264   &   --23.70 & --12.00$\pm0.5$  \\
          3C~66   &   --23.74 & --12.04$\pm0.5$  \\
       3C~272.1   &   --23.74 & --12.05$\pm0.5$  \\
         3C~315   &   --23.82 & --12.12$\pm0.5$  \\
         3C~338   &   --23.98 & --12.28$\pm0.5$  \\
         3C~293   &   --24.00 & --12.30$\pm0.5$  \\
          3C~29   &   --24.03 & --12.33$\pm0.5$  \\
          3C~31   &   --24.04 & --12.34$\pm0.5$  \\
         3C~310   &   --24.10 & --12.40$\pm0.5$  \\
         3C~296   &   --24.11 & --12.41$\pm0.5$  \\
          3C~89   &   --24.31 & --12.61$\pm0.5$  \\
          3C~75   &   --24.41 & --12.71$\pm0.5$  \\
         3C~449   &   --24.43 & --12.73$\pm0.5$  \\
         3C~288   &   --24.52 & --12.82$\pm0.5$  \\
         3C~305   &   --24.53 & --12.83$\pm0.5$  \\
        3C~83.1   &   --24.68 & --12.98$\pm0.5$  \\
         3C~424   &   --24.74 & --13.05$\pm0.5$  \\
         3C~438   &   --24.77 & --13.07$\pm0.5$  \\
         3C~386   &   --24.77 & --13.07$\pm0.5$  \\
       3C~277.3   &   --24.91 & --13.21$\pm0.5$  \\
         3C~348   &   --25.00 & --13.30$\pm0.5$  \\
         3C~433   &   --25.30 & --13.60$\pm0.5$  \\
         3C~442   &   --25.70 & --14.00$\pm0.5$  \\
\hline
\end{tabular}
\caption{
Radio core fluxes at 5 GHz for 3CR FR~I and predicted gamma--ray fluxes. 
See the text for the assumptions.}
\end{center}
\end{table}

\section{Discussion}

We have explored the radiative and dynamical consequences
of a structured jet, in which a slow jet layer is cospatial to
a fast jet spine. 
The motivations for this study are primarily observational,
since radio data have recently found slow or null proper
motion for the parces scale radio knots in TeV emitting BL Lacs.
In addition, detailed VLBI radio maps show hints of a limb
brightening for the jet in Mkn 501.

The jet may be born as structured, or it may be born with equal layer/spine 
velocities, with the layer being decelerated in the first
$\sim 10^{17}$ cm from the central black hole.
As long as the layer is dissipative, our results are independent 
of the mechanism producing a spine+layer structure.
However, we note that the accretion disk in low power
BL Lacs could be characterized by a low--efficiency accretion
mode, and one of the proposed accretion disk solution in this
case is the ADIOS structure (Blandford \& Begelman 1999),
predicting that a sizeable fraction of the accreting mass
at large radii, instead to inward spiralling, leaves the
disk in the form of an outflow.
In this case, in the close vicinity of the
black hole, the ambient medium should be relatively
dense, favoring the interaction of the jet walls 
with the external medium.
This may cause the formation of the slow layer, and at the same
time this interaction may be the primary
cause for the transformation of the kinetic energy of the layer
into random energy and then radiation.

A low radiative efficiency accretion disk could also explain why
the broad line emission is small in BL Lac objects and in FR I
radiogalaxies. 
Further evidence of a change in the accretion mode between
FR I and FR II radiogalaxies is provided 
by the ``dividing line" between these
two classes of sources in the radio-luminosity--host galaxy
optical luminosity plane, as discussed by Ghisellini \& Celotti (2001).

If indeed the jet forms (or is born with) a layer+spine structure, then
there is a radiative interplay or feedback between the two parts:
each component would see an enhanced radiation field coming from the
other component.
This would inevitably boost the inverse Compton radiation
with respect to a completely homogeneous jet.

It is also quite clear that, for an observed Compton to synchrotron
power ratio, the fact that the radiation field is enhanced also
implies an increase of the magnetic field, with respect to
an homogeneous source.
This can solve an otherwise puzzling characteristic of
HBL in general and TeV BL Lacs in particular: fitted with
an homogeneous SSC model, they turn out to have very small 
magnetic fields, which are more under--equipartition 
(with the emitting particles) than in other blazars.
Here we can fit the observed spectra equally well
(not surprisingly, given that the free parameters are more
than in the homogeneous SSC model) with equipartition
magnetic fields.

Another important consequence of having greater magnetic fields
is that a smaller number of electrons can produce the observed SED.
This means that the global energetic demand of the jet is reduced
with respect to a homogeneous SSC model.

The fact that the dominant inverse Compton radiation is through
scattering with ``external" photons implies that the emission
is highly anisotropic also in the comoving frame of the spine.
To conserve momentum, the emitting spine must recoil and
therefore decelerate.
This is a manifestation of the ``Compton rocket" effect,
studied in the early eighties (see e.g. O'Dell 1981) 
as the means to radiatively accelerate jets. 
Somewhat ironically, we have shown here that this process
can be important for the opposite reason.
There is a precise link between TeV emission and deceleration,
due to two reasons:
i) TeV emitting BL Lacs have the least powerful jets, and yet
they move with bulk Lorentz factor equal or greater than the ones
of other blazars;
ii) to produce a significant TeV radiation, the mean energy of the emitting
electrons must be large. Then in these sources we have
$\langle\gamma\rangle m_e \sim m_p$: the power carried by the jet
in the form of protons and electrons is similar.
We then propose that the jet deceleration at small (sub--pc) scales
is more efficient in low power jets emitting high energy radiation.
Consider also that the initially fastest sources are the ones suffering
the most severe Compton rocket effect.

Also the inverse Compton emission from the layer is enhanced
by the extra seed photons coming from the spine.
This could be the reason why also radiogalaxies are relatively
strong $\gamma$--ray emitters.
If this is the reason, then the layer and the spine must be cospatial,
and therefore the $\gamma$--ray emitting layer must be located
at $\sim$100 Schwarzchild radii, as the spine.
The $\gamma$--ray flux observed in radiogalaxies
should then be variable, with timescales
of the order of $t_{\rm var}\sim (R/c)/\delta_l\sim$ a day or less,
very similar (albeit somewhat longer, due to the smaller Doppler
factor) to the typical variability timescale in blazars.
We propose this as a crucial test for our scenario.
Note that Cen A is already known to vary with short timescales
(0.5--4 days) in the $\gamma$--ray band (Kinzer et al. 1995; 
Steinle et al. 1998).

A few radiogalaxies have been already detected at high energies.
The SED of their nuclear emission shows the characteristic double peak
characteristic of blazars, and like these sources they can
be ``Compton dominated" (namely, the high energy component
is more luminous than the synchrotron component).
These general features are quite easily explained 
in the layer+spine scenario.
We have then tried to predict which are the best candidate
radiogalaxies to be detected by the future GLAST mission,
assuming a twenty--fold increase in sensitivity with respect
to EGRET. 
As a result, more than a dozen FR I radiogalaxies
should be detectable by GLAST, if their radio to
100 MeV flux ratio is similar to the one of the three
objects already identified by EGRET and by TeV Cherenkov
telescopes.




\begin{thebibliography}{}
\bibitem[]{} Aharonian F., Akhperjanian A., Beilicke M. et al., 2003, A\&A, 403 L1
\bibitem[]{} Blandford R.D. \& Begelman M.C., 1999, MNRAS, 303, L1
\bibitem[]{} Bodo G., Rossi P., Mignone A., Massaglia S. \& Ferrari A.,
              2003, NewAR, 47, 557
\bibitem[]{} Cassaro P., Stanghellini C., Bondi M., Dallacasa D., 
\bibitem[]{} Chiaberge M., Capetti A., \& Celotti A., 1999, A\&A, 349, 77
\bibitem[]{} Chiaberge M., Celotti A., Capetti A. \& Ghisellini G.,
          2000, A\&A, 358, 104
\bibitem[]{} Chiaberge M., Gilli, R., Capetti A., Macchetto F.D.,
          2003, ApJ, 97, 166 
\bibitem[]{} Chiaberge M., Capetti A. \& Celotti A.,  2001, MNRAS, 324, L33 
\bibitem[]{} Costamante L. \& Ghisellini G., 2002, A\&A, 384, 56
\bibitem[]{} Dermer C.D., 1995, ApJ, 446, L63
\bibitem[]{} Djannati-Atai A., Piron F., Barrau A. et al., A\&A, 350, 17
\bibitem[]{} Donato D., Ghisellini G., Tagliaferri G. \& Fossati G., 2001,
   A\&A, 375, 739
\bibitem[]{} Dondi L. \& Ghisellini G., 1995, MNRAS, 273, 583
\bibitem[]{} Edwards P.G. \& Piner B.G., 2002, ApJ, 579, L70
\bibitem[]{} Ferrari A., 1998, ARA\&A, 36, 539
\bibitem[]{} Fichtel C.E., Bertsch D.L., Chiang J. et al., 1994, ApJS, 94, 551
\bibitem[]{} Fossati G., Maraschi L., Celotti A., Comastri A. \& Ghisellini G.,
     1998, MNRAS, 299, 433
\bibitem[]{} Georganopoulos M. \& Kazanas D., 2003, ApJ, 594, L27 (GK03)
\bibitem[]{} Ghisellini G., 1999, Astron. Nachrichten, 320, 232.
\bibitem[]{} Ghisellini G. \& Celotti A., 2001, A\&A, 379, L1 
\bibitem[]{} Ghisellini G., Celotti A., Fossati G., Maraschi L. \& Comastri A.,
        1998, MNRAS, 301, 451 
\bibitem[]{} Ghisellini G., Celotti A. \& Costamante L., 2002, A\&A, 386, 833 
\bibitem[]{} Giommi P. \& Padovani P., 1995, MNRAS, 277, 1477
\bibitem[]{} Giovannini G., Feretti L., Gregorini L., \& Parma P., 
          1988, A\&A, 199, 73
\bibitem[]{} Giovannini G., Taylor G.B., Arbizzani E. et al., 1999, ApJ, 522, 101
\bibitem[]{} Giroletti M., Giovannini G., Feretti L. et al., 2004, ApJ, 600, 127
\bibitem[]{} Gliozzi M., Sambruna R.M., Brandt W.N., Mushotzky R. \& Eracleous M., 
          2004, A\&A, 413, 139
\bibitem[]{} Guetta D., Ghisellini G., Lazzati D. \& Celotti A., 2004, A\&A, 421, 877
\bibitem[]{} Hartman R.C., Bertsch S.D., Bloom S.D. et al., 1999, ApJS, 123, 79
\bibitem[]{} Henri, G.~\& Pelletier, G.\ 1991, ApJ, 383, L7 
\bibitem[]{} Katarzynski K., Sol H. \& Kus A., 2003, A\&A, 410, 101
\bibitem[]{} Kino M., Takahara F. \& Kusunose M., 2002, ApJ, 564, 97
\bibitem[]{} Kinzer R.L., Johnson W.N., Dermer C.D. et al., ApJ, 449, 105
\bibitem[]{} Komissarov S.S., 1990, SvA Letters, 16, 284 
\bibitem[]{} Konopelko K., Mastichiadis A., Kirk J., De Jager O.C. \& Stecker F.W.,
              2003, ApJ, 597, 851
\bibitem[]{} Krawczynski H., Coppi P.S. \& Aharonian F., 2002, MNRAS, 336, 721
\bibitem[]{} Krawczynski H., Hughes S.B., Horan, D. et al., 2004, ApJ, 601, 151
\bibitem[]{} Laing R.A., 1993, in: Astrophysical jets, Space Telescope Sci. Inst.
             Symp. 6, eds: Burgarella D., Livio M. \& O'Dea C.P., 
             Cambridge Univ. Press, p. 95
\bibitem[]{} Laing R.A., Riley J.M. \& Longair M.S., 1983, MNRAS, 204, 151
\bibitem[]{} Morganti R., Fosbury R.A.E., Hook R.N., Robinson A. \& Tsvetanov Z.
    1992, MNRAS, 256, P1
\bibitem[]{} Marshall H.L., Miller B.P., Davis D.S., Perlman E.S., Wise M., Canizares C.R. \&
   Harris D.E. 2002, ApJ, 564, 683
\bibitem[]{} Mukherjee R., Halpern J., Mirabal N. \& Gotthelf E.V., 2002, ApJ, 574, 693
\bibitem[]{} O'Dell S.L., 1981, ApJ, 243, L147
\bibitem[]{} Owen F.N., Hardee P.E. \& Cornwell T.J., 1989, ApJ 340, 698
\bibitem[]{} Pian E., Vacanti G., Tagliaferri G. et al., 1998, ApJ, 492, L17
\bibitem[]{} Piner B.G. \& Edwards P.G., 2004, ApJ, 600, 115
\bibitem[]{} Rawlings S.G. \& Saunders R.D.E., 1991, Nature, 349, 138
\bibitem[]{} Reimer A., Protheroe R.J. \& Donea A.--C., 2004, A\&A 419, 89
\bibitem[]{} Rybicki G.B. \& Lightman A.P., 1979, Radiative Processes in Astrophysics,
   New Youk, Wiley Interscience
\bibitem[]{} Rossi E.M., Lazzati D. \& Rees, M.J., 2002, MNRAS. 332, 945
\bibitem[]{} Sikora M., Sol H., Begelman M.C. \& Madejski G.M., 
               1996, MNRAS, 280, 781
\bibitem[]{} Spada M., Ghisellini G., Lazzati D \& Celotti A., 2001, MNRAS, 325, 1559 
\bibitem[]{} Stawarz L. Sikora M., Ostrowski M. 2003, ApJ, 597, 186
\bibitem[]{} Steinle H., Bennet K., Bloemen H. et al., 1998, A\&A, 330, 97
\bibitem[]{} Swain M.R., Bridle A.H. \& Baum S.A., 1998, ApJ, 507, L29
\bibitem[]{} Tavecchio F., Maraschi L. \& Ghisellini G., 1998. ApJ, 509, 608 
\bibitem[]{} Tavecchio F., Maraschi L., Pian E. et al., 2001, ApJ, 554, 725
\bibitem[]{} Zhang B. \& Meszaros P., 2002, ApJ, 571, 876

\end{thebibliography}
\end{document}